\documentclass[preprint]{aastex}

\usepackage{emulateapj5}
\usepackage{apjfonts}

\newcommand{\Teff}{\mbox{$T_{\rm eff}$}}
\newcommand{\Dnu}{\mbox{$\Delta \nu$}}
\newcommand{\acena}{\mbox{$\alpha$~Cen~A}}
\newcommand{\acenb}{\mbox{$\alpha$~Cen~B}}

\newcommand{\bvir}{\mbox{$\beta$~Vir}}
\newcommand{\nuind}{\mbox{$\nu$~Ind}}
\newcommand{\muara}{\mbox{$\mu$~Ara}}
\newcommand{\cms}{\mbox{cm\,s$^{-1}$}}
\newcommand{\eboo}{\mbox{$\eta$~Boo}}
\newcommand{\ms}{\mbox{m\,s$^{-1}$}}
\newcommand{\muHz}{\mbox{$\mu$Hz}}
\newcommand{\new}[1]{{\bf #1}}
\renewcommand{\new}[1]{{\relax #1}}
\newcommand{\half}{{\textstyle\frac{1}{2}}}

\slugcomment{To appear in ApJ}

\shorttitle{Oscillations in $\nu$ Indi}
\shortauthors{Bedding et al.}

\begin{document}

\title{Solar-like oscillations in the metal-poor subgiant $\nu$~Indi:
  constraining the mass and age using asteroseismology}

\author{
Timothy R. Bedding,\altaffilmark{1}
R.~Paul~Butler,\altaffilmark{2}
Fabien~Carrier,\altaffilmark{3}
Francois~Bouchy,\altaffilmark{3,4}
Brendon~J.~Brewer,\altaffilmark{1}
Patrick~Eggenberger,\altaffilmark{3}
Frank~Grundahl,\altaffilmark{5}
Hans~Kjeldsen,\altaffilmark{5}
Chris~McCarthy,\altaffilmark{2}
Tine~Bj{\o}rn~Nielsen,\altaffilmark{5}
Alon~Retter,\altaffilmark{1,6}
Christopher~G.~Tinney\altaffilmark{7}
}

\altaffiltext{1}{School of Physics A28, University of Sydney, NSW 2006,
Australia; bedding@physics.usyd.edu.au; brewer@physics.usyd.edu.au}

\altaffiltext{2}{Carnegie Institution of Washington,
Department of Terrestrial Magnetism, 5241 Broad Branch Road NW, Washington,
DC 20015-1305; paul@dtm.ciw.edu, chris@dtm.ciw.edu}

\altaffiltext{3}{Observatoire de Gen\`eve, Ch.~des Maillettes 51, CH-1290
Sauverny, Switzerland; fabien.carrier@obs.unige.ch;
francois.bouchy@obs.unige.ch; patrick.eggenberger@obs.unige.ch}

\altaffiltext{4}{Laboratoire d'Astrophysique de Marseille, Traverse du
Siphon, BP 8, 13376 Marseille Cedex 12, France}

\altaffiltext{5}{Department of Physics and Astronomy, University of Aarhus,
DK-8000 Aarhus C, Denmark; hans@phys.au.dk, fgj@phys.au.dk, tbn@phys.au.dk}

\altaffiltext{6}{Department of Astronomy and Astrophysics, Pennsylvania
State University, 525 Davey Lab, University Park, PA 16802-6305;
retter@astro.psu.edu}

\altaffiltext{7}{Anglo-Australian Observatory, P.O.\,Box 296, Epping, NSW
1710, Australia; cgt@aaoepp.aao.gov.au}

\begin{abstract} 
Asteroseismology is a powerful method for determining fundamental
properties of stars.  We report the first application to a metal-poor
object, namely the subgiant star \nuind.  We measured precise velocities
from two sites, allowing us to detect oscillations and infer a large
frequency separation of $\Delta \nu = 24.25 \pm 0.25 \mu$Hz.  Combining
this value with the location of the star in the H-R diagram and comparing
with standard evolutionary models, we were able to place constraints on the
stellar parameters.  In particular, our results indicate that \nuind{} has
a low mass ($0.85\pm0.04\,M_\sun$) and is at least 9\,Gyr old.
\end{abstract}

%% Asteroseismology is a powerful method for determining fundamental
%% properties of stars.  We report the first application to a metal-poor
%% object, namely the subgiant star nu Ind.  We measured precise velocities
%% from two sites, allowing us to detect oscillations and infer a large
%% frequency separation of 24.25 +/- 0.25 microHz.  Combining this value with
%% the location of the star in the H-R diagram and comparing with standard
%% evolutionary models, we were able to place constraints on the stellar
%% parameters.  In particular, our results indicate that nu Ind has a low mass
%% (0.85 +/- 0.04 M_sun) and is at least 9 Gyr old.


\section{Introduction}

Asteroseismology is a powerful method for determining fundamental
properties of stars.  This is because oscillation frequencies give strong
constraints on the internal structure that are independent of classical
observations.  Observations of solar-like oscillations are accumulating
rapidly, and measurement have recently be reported for several
main-sequence and subgiant stars, including
\acena{} \citep{B+C2002,BKB2004}, 
\acenb{} \citep{C+B2003,KBB2005},
\bvir{} \citep{MLA2004b,CEDAl2005},
\muara{} \citep{BBS2005}, 
HD~49933 \citep{MBC2005} and
\eboo{}  \citep{KBB2003,CEB2005,GKR2005}.
All of these stars have solar metallicity or greater.  Asteroseismology is
particularly useful for constraining the evolutionary status of stars with
low metallicity \citep[e.g.][]{DACDiM2005}.  Here, we report the first
oscillation measurements of a metal-poor star.

We have observed the subgiant \nuind{} (HR~8515; HD 211998; HIP 110618),
whose iron abundance is only 3\% of solar.  We adopt the following stellar
parameters: $V=5.28$, $\Teff = 5300 \pm 100$\,K, $L=6.2\,L_\sun$ and
$\mbox{[Fe/H]}=-1.4\pm0.1$ \citep{NHS97,GSC2000}.  Note that the Bright
Star Catalogue \citep{Hoffleit} incorrectly lists this star as a binary
with spectral types A3V:+F9V.  In fact, \nuind{} is a single star with
spectral type~G0 \citep{L+McW86}.

\section{Velocity observations and power spectra}

We observed \nuind{} in August 2002 from two sites.  
At Siding Spring Observatory in Australia we used UCLES (University College
London Echelle Spectrograph) with the 3.9-m Anglo-Australian Telescope
(AAT).  An iodine absorption cell was used to provide a stable wavelength
reference, with the same setup that we have previously used with this
spectrograph \citep{BBK2004}.
At the European Southern Observatory on La Silla in Chile we used the
CORALIE spectrograph with the 1.2-m Swiss telescope.  A thorium emission
lamp was used to provide a stable wavelength reference, and the velocities
were processed using the method described by \citet{BPQ2001}.

With UCLES we obtained 680 spectra of \nuind, with typical exposure times
of 300\,s (but sometimes as short as 200\,s in the best conditions) and a
dead time between exposures of 61\,s.  With CORALIE we obtained 521
spectra, with typical exposure times of 360\,s and a dead time between
exposures of 128\,s.

\begin{figure*}
\epsscale{0.9}
\plotone{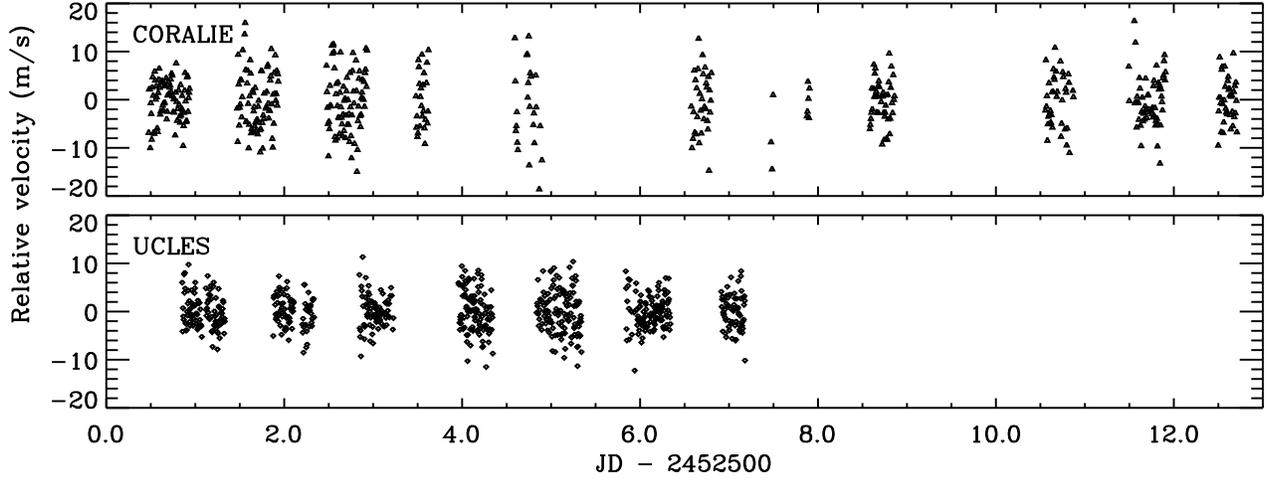}
\caption[]{\label{fig.series} Time series of velocity measurements of
\nuind{} from the UCLES and CORALIE spectrographs.  The mean of each night
has been subtracted. }
\end{figure*}

The resulting velocities, with nightly means subtracted, are shown in
Fig.~\ref{fig.series}.  As can be seen, the weather was good in Australia
but poor in Chile (we were allocated seven nights with UCLES and 14 with
CORALIE).

\begin{figure*}
\epsscale{0.9}
\plotone{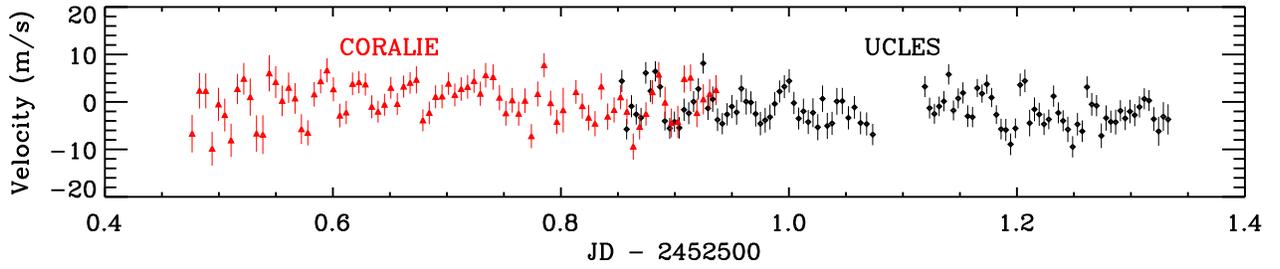}
\caption[]{\label{fig.best} Velocities with 1-$\sigma$ error bars for the
  first night at each site.  The 50-minute oscillations are clearly seen.
  }
\end{figure*}

Most of the scatter in the velocities, especially for UCLES, is due to
oscillations.  Figure~\ref{fig.best} shows a close-up of the first night.
Oscillations with a period of about 50\,min and variable amplitude (due to
beating between modes) are visible here and throughout the time series.  We
also see good agreement between the two instruments, within measurement
uncertainties, during the overlap (although we should note that the
difference in absolute velocities is not known and has been adjusted to
give the best fit).

\begin{figure*}
\epsscale{0.9}
\plotone{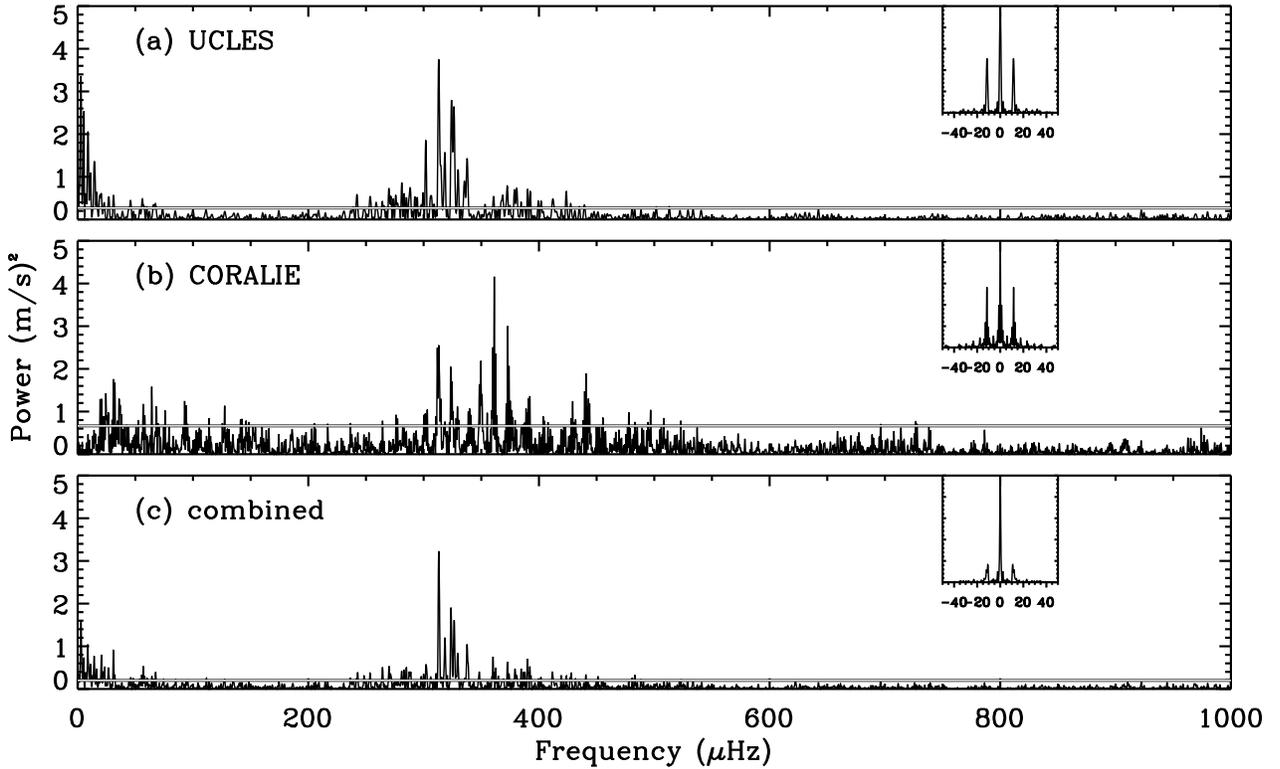}
\caption[]{\label{fig.power} Power spectra of velocity measurements of
\nuind{} from the two time series and from the combined data.  \new{The
horizontal lines show $(3\sigma)^2$, where $\sigma$ is the noise at high
frequencies in each amplitude spectrum.}  The inset in each panel shows the
spectral window. }
\end{figure*}

Our analysis of these velocities follows the method that we developed for
\acena{} \citep{BBK2004} and \acenb{} \citep{KBB2005}.  We have used the
measurement uncertainties, $\sigma_i$, as weights in calculating the power
spectrum (according to $w_i = 1/\sigma_i^2$), but modified some of the
weights to account for a small fraction of bad data points.  In this case,
4 data points from UCLES and 24 from CORALIE needed to be down-weighted.
The power spectra of the individual time series and of the combination are
shown in Fig.~\ref{fig.power}.  The differences between the three panels
can be attributed to the effects of beating between modes.

The low metallicity of \nuind{} means that the lines in its spectrum are
fewer and weaker than for stars of solar metallicity.  This, together with
its relative faintness ($V=5.3$), means that the Doppler precision for
\nuind{} is poorer than for other stars observed with the same instruments,
such as \acena{} and~B.  We measured the average noise in the amplitude
spectrum of \nuind{} at frequencies above the stellar signal (0.6--1\,mHz)
to be 17.3\,\cms{} for UCLES, 27.2\,\cms{} for CORALIE and 14.9\,\cms{} for
the combined data.  Using these values, we calculated the noise per minute
of observing time to be 5.9\,\ms{} for UCLES and 9.5\,\ms{} for CORALIE,
where the difference is due to a combination of factors, primarily the
telescope aperture but also including spectrograph design, sky conditions
and observing duty cycle.

The inset in each panel of Fig.~\ref{fig.power} shows the spectral window
(the response to a single pure sinusoid).  For the single-site data we see
sidelobes at $\pm 11.6$\,\muHz{} that are very strong (51\% in power for
UCLES and 57\% for CORALIE).  These are due to daytime gaps in the
observing window.  When the data are combined, the sidelobes are
drastically reduced (to 16\% in power) and are also slightly shifted,
occurring at $\pm 10.8$\,\muHz.  In the cases of \acena{} and B, we
generated power spectra in which the weights were adjusted on a
night-by-night basis in order to minimize the sidelobes.  We have not done
that for \nuind{} because the sidelobes for the two-site data are already
quite low.

\section{Large frequency separation}    \label{sec.Dnu}

Mode frequencies for low-degree p-mode oscillations are approximated
reasonably well by a regular series of peaks:
\begin{equation}
  \nu_{n,l} = \Dnu{} (n + \half l + \epsilon) - l(l+1) D_0.
        \label{eq.asymptotic}
\end{equation}
Here $n$ (the radial order) and $l$ (the angular degree) are integers,
$\Dnu{}$ (the large separation) depends on the sound travel time across the
whole star, $D_0$ is sensitive to the sound speed near the core and
$\epsilon$ is sensitive to the surface layers.  See \citet{ChD2004} for a
recent review of the theory of solar-like oscillations.

\begin{figure*}
\epsscale{0.5}
\plotone{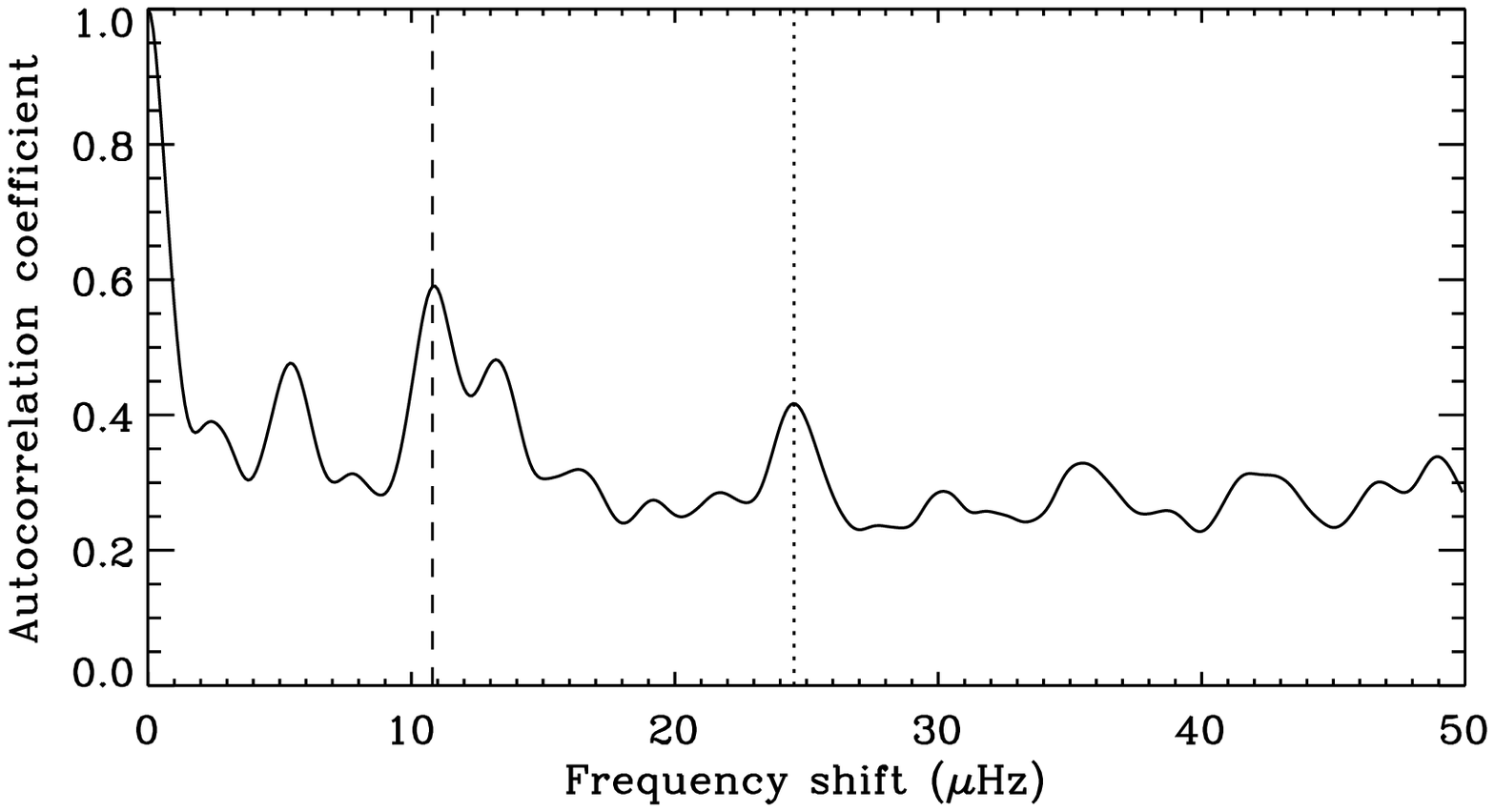}\\
\bigskip
\plotone{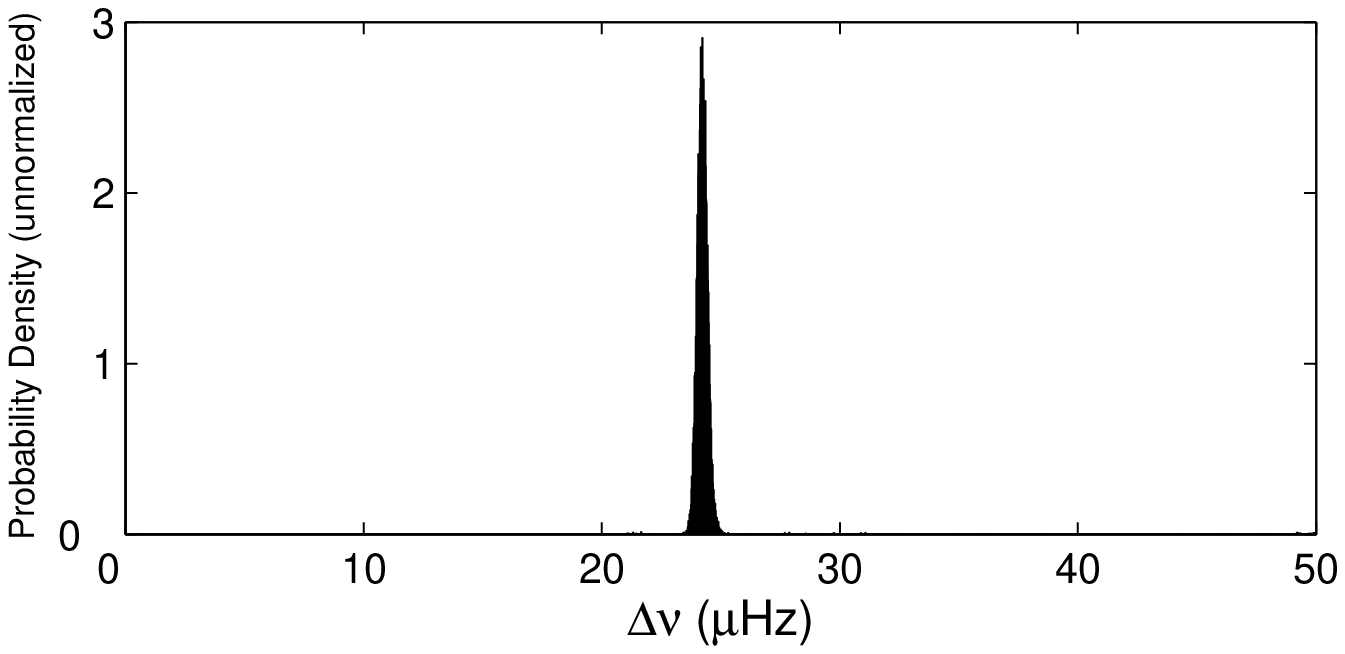}
\caption[]{\label{fig.acorr} Measurements of the large separation of
\nuind.  Upper panel: the autocorrelation of the power spectrum, with peaks
at 24.5\,\muHz{} (dotted line) from the large separation and at
10.8\,\muHz{} (dashed line) from the daily sidelobes.
\label{fig.bayes} Lower panel: probability distribution for the large
separation calculated using Bayesian methods (see text).  }

\end{figure*}

The large separation, $\Dnu{}$, is proportional to the square root of the
mean density of the star.  Scaling from the Sun, for which $\Dnu{} =
135\,\muHz$, we expect \nuind{} to have a large separation of about
25\,\muHz.  In order to search for a regular series of peaks from which to
measure the large separation, we calculated the autocorrelation function of
the power spectrum.  The result is shown in Fig.~\ref{fig.acorr} and the
peak at 10.8\,\muHz{} (dashed line) is the main sidelobe caused by the
daily gaps (see above).  We interpret the peak at 24.5\,\muHz{} as due to
the large separation, with smaller peaks at $\pm 10.8\,\muHz$ from this
being due to daily aliases \new{(the peak at 13\,\muHz{} also coincides
with half the large separation).  The peak at 5\,\muHz{} is not easily
explained by the regular p-mode structure and may reflect departures of a
few modes from the asymptotic relation in equation~\ref{eq.asymptotic}.}

As a check of this result, we also measured the large separation directly
from the light curve data using Bayesian methods (see \citealt{Gre2005} for
a good introduction to Bayesian analysis).  We modelled the light curve as
a sum of sinusoids, where the number of sinusoids, together with their
amplitudes, phases and frequencies, were treated as unknowns.  An advantage
of this approach is that the amplitudes and phases can be integrated out of
the problem at the beginning \citep{Bre88}, so that we only need to fit the
frequencies and their number.  To determine the large separation, we chose
the prior distribution for the frequencies to be periodic, with period
$\frac{1}{2} \Delta \nu$ (considered unknown).  We fixed the central
frequency of this periodic comb to be the highest peak in the power
spectrum (313.14\,\muHz).  We then calculated the posterior distribution
for $\Delta \nu$, which is shown in the lower panel of
Fig.~\ref{fig.bayes}.  There is a single strong peak with a value of
$\Delta \nu = 24.25 \pm 0.25 \mu$Hz, in agreement with the peak of the
autocorrelation.  More details of this method, which is similar to the
approach used by \citet{Bre2003} and appears to be very promising for this
type of analysis, will be presented separately (B. Brewer et al., in
prep.).  

\section{Constraints on the stellar parameters}

\begin{figure*}
\epsscale{0.4} 
\plotone{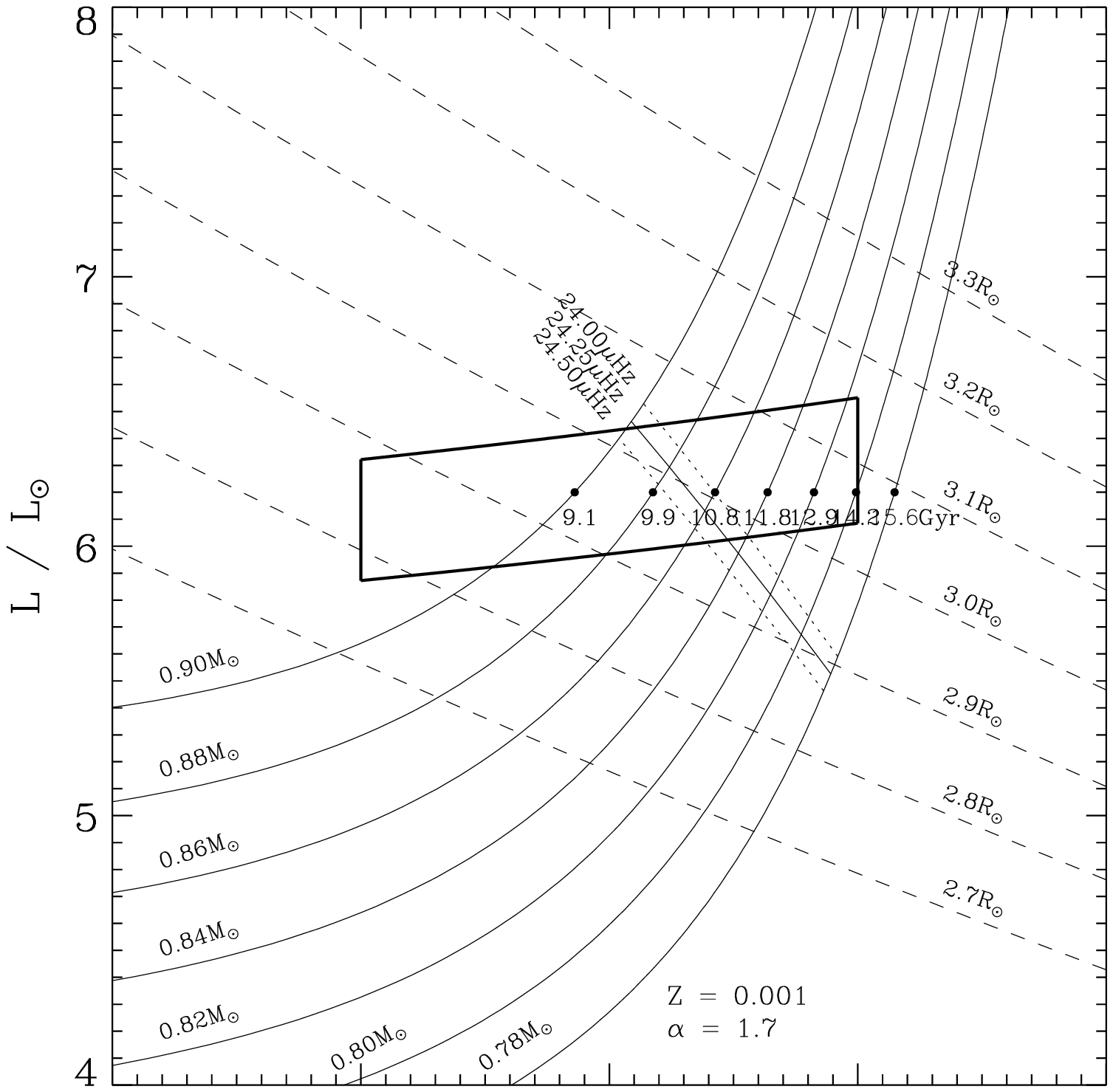}\\ \plotone{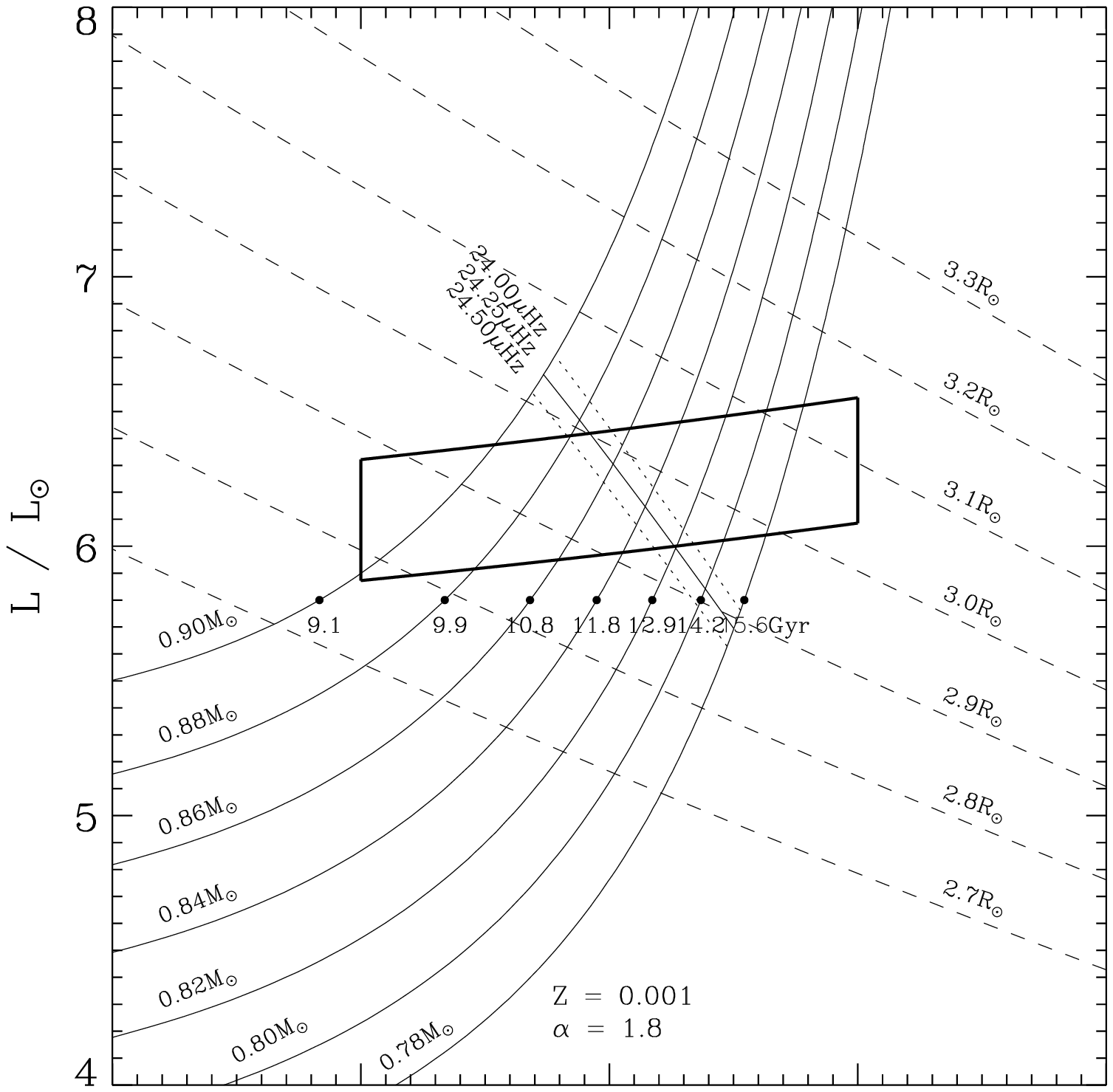}\\ \plotone{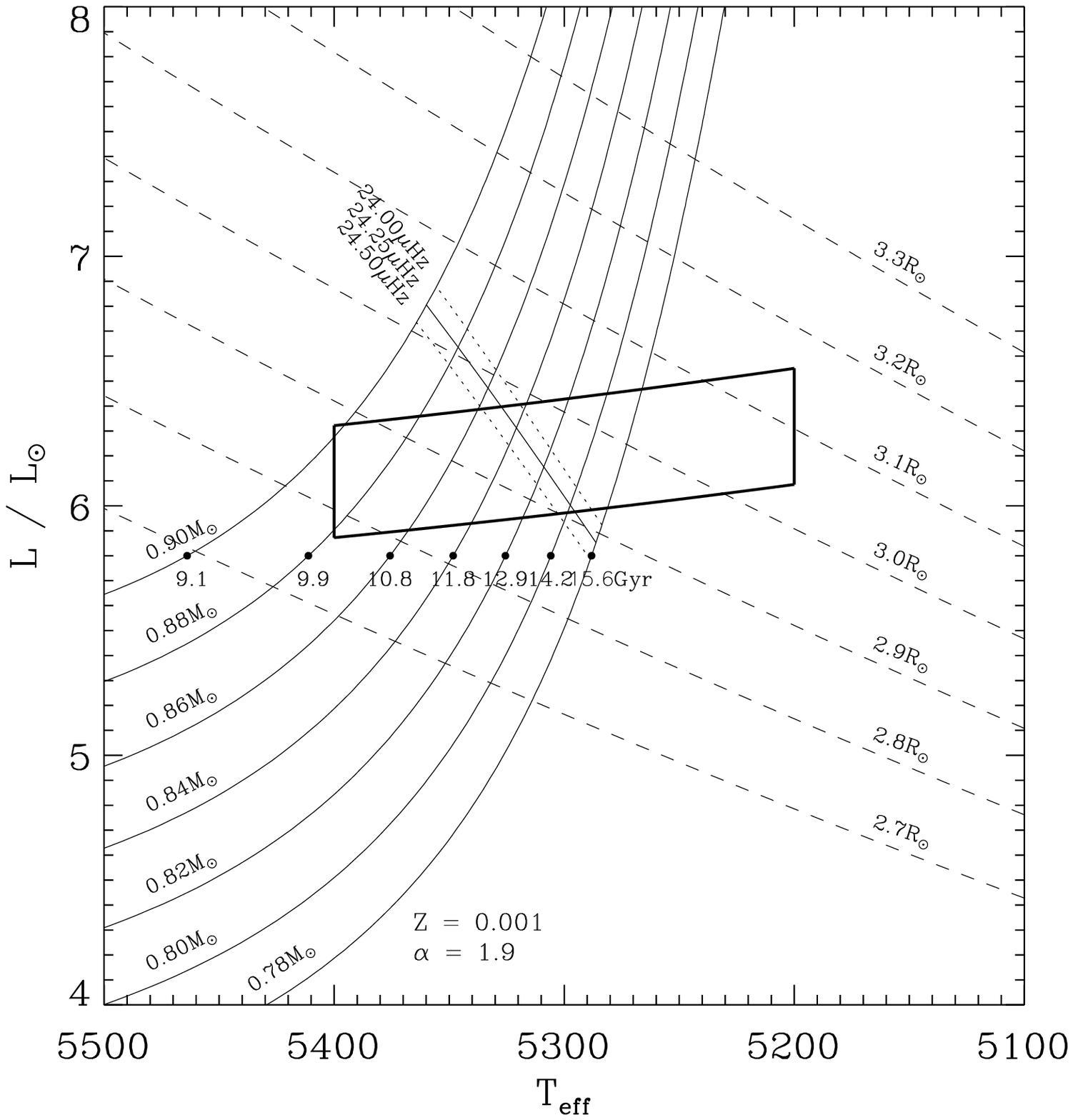}
\caption[]{\label{fig.hr} H-R diagrams in which the box shows the position
of \nuind{} from classical observations and the diagonal dashed lines are
loci of constant radius.  The curved lines are evolutionary tracks for
models with $Z=0.001$ and a range of masses, with the three panels
differing in the value of the mixing-length parameter,~$\alpha$.  The
diagonal lines are loci of constant $\Dnu{}$, calculated from the mean
densities of the models by scaling from the Sun.
}
\end{figure*}

What can we learn about \nuind{} from our measurement of \Dnu?
Figure~\ref{fig.hr} shows the location of the star in the H-R diagram,
together with some theoretical evolutionary tracks.  The box, which is the
same in each panel, shows the observed position of \nuind{} from classical
measurements.  The value for \Teff{} ($5300 \pm 100$\,K) is the mean of
published photometric estimates \citep{NHS97,GSC2000}, where we note the
large uncertainty in the effective temperature scale for metal-poor stars.
The luminosity is based on the Hipparcos parallax ($34.6 \pm 0.6$\,mas),
with bolometric corrections from \citet{AAMR99}.  Note that the bolometric
correction is a function of effective temperature, hence the slope of the
box.  The diagonal dashed lines, which are also the same in each panel, are
loci of constant radius, calculated from $L\propto R^2\Teff^4$.  We can
immediately see that a measurement of the radius from interferometry would
be valuable in constraining the location of the star in the H-R diagram, as
has already been shown for other stars
\citep{KTS2003,PTG2003,KTM2004,TKP2005}.

The curved lines in Fig.~\ref{fig.hr} are evolutionary tracks for a range
of masses, using model calculations similar to those by \citet{ChD82}.  We
used a metallicity of $Z=0.001$ (with hydrogen and helium mass fractions of
$X=0.75$ and $Y=0.249$, respectively) and the three panels differ in the
adopted value of the mixing-length parameter ($\alpha = 1.7,$ 1.8 and~1.9),
where the solar value is $\alpha_\sun=1.83$.  The relatively rapid
evolution in this subgiant phase means each track can be described by a
single age, as shown in the figure.  Finally, the diagonal lines are
contours of constant $\Dnu{}$.  We calculated these from the evolutionary
models by scaling from the Sun (since $\Dnu{}$ is proportional to the
square root of the mean density).

\begin{figure*}
\epsscale{0.5}
\plotone{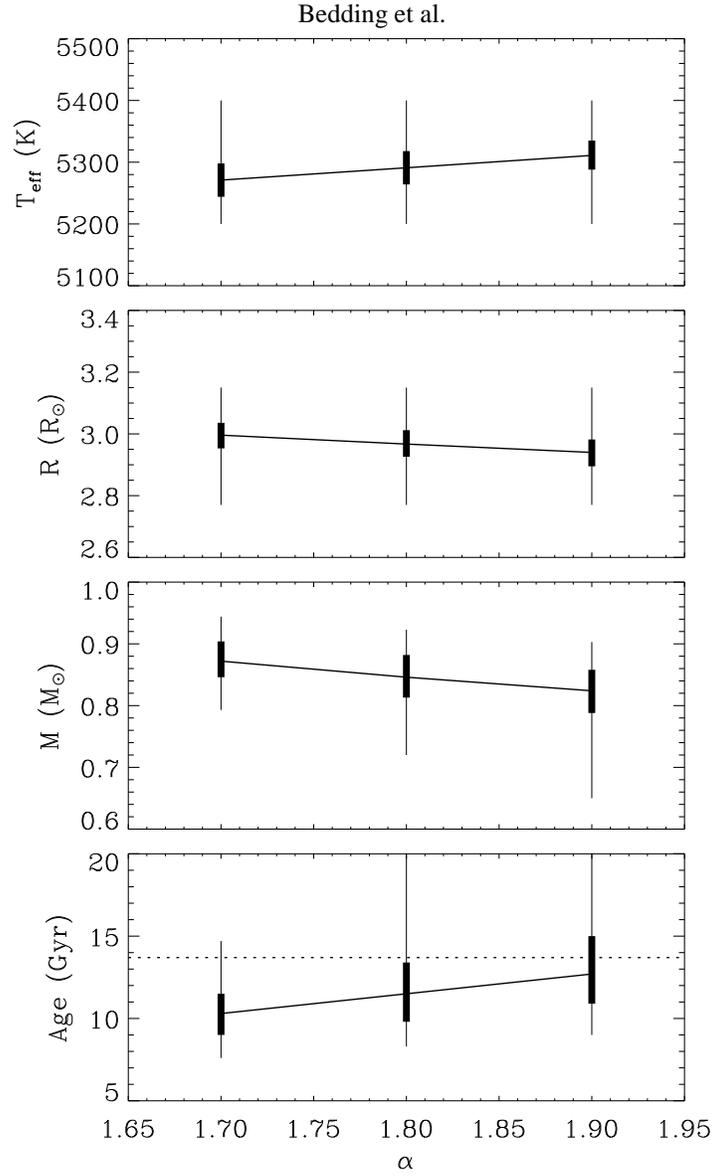}
\caption[]{\label{fig.params} Parameters of \nuind{} based on observations
and models, as a function of the mixing-length parameter.  The thin error
bars show the range of each parameter based on classical measurements alone
($L$ and $\Teff$), while the thick bars include the constraint provided by
our measurement of $\Dnu$.  The dashed line at an age of 13.7\,Gyr
indicates the upper limit set by age of the universe \citep{SVP2003}.  }
\end{figure*}

We see that our measurement of $\Dnu{}$ significantly constrains the
parameters of \nuind.  This is quantified in Fig.~\ref{fig.params}, in
which the thin error bars show the range of each parameter based on
classical measurements alone ($L$ and $\Teff$), while the thick bars show
the situation after we have added the constraint provided by our
measurement of $\Dnu$.  Including this constraint reduces the uncertainty
in both effective temperature and radius, for a given value of $\alpha$, by
a factor of four.

What can we say about the mass of \nuind?  Even with no constraints from
seismology, the requirement that \nuind{} be younger than the universe
(13.7\,Gyr; \citealt{SVP2003}) sets a lower limit of $0.81\,M_\sun$.  A
star of lower mass would not have had time to evolve this far.  Note that
this limit, which we derived from the tracks in Fig.~\ref{fig.params}, is
essentially independent of the mixing length.  This is because the value of
$\alpha$ has little effect on the fusion rate in the core, and hence on the
time for a star of given mass to leave the main sequence and enter the
subgiant phase.  Meanwhile, an upper limit on the mass is obtained from our
measurement of $\Dnu$, provided we also set a lower limit on $\alpha$
(third panel of Fig.~\ref{fig.params}).  Adopting a plausible lower limit
of $\alpha\ge1.7$ gives an upper limit for the mass of $0.89\,M_\sun$.

We can also set interesting limits on the age of \nuind.  From the bottom
panel of Fig.~\ref{fig.params}, and again setting $\alpha>1.7$ as a
plausible limit, we see that the age must be at least 9\,Gyr.  This
confirms that \nuind{} is, indeed, very old and must have been formed very
early in the history of the Galaxy.  The final results of our analysis are
summarized in Table~\ref{tab.params}, where we list our best estimates for
the parameters of \nuind, assuming that $Z=0.001$ and $\alpha =
1.8\pm0.1$.  

With the data currently available, what constraints can we set on the
mixing length?  The dashed line at an age of 13.7\,Gyr in
Fig.~\ref{fig.params} indicates the upper limit set by age of the universe.
This indicates that the mixing length cannot be greater than 2.1.  We
expect much stronger constraints on $\alpha$, and also on the other stellar
parameters, to come from the individual oscillation frequencies.  The
extraction of these frequencies and a comparison with theoretical models is
deferred to a future paper (F. Carrier et al., in prep.).  An accurate
measurement of the radius using interferometry would also be extremely
valuable.

%%%%%%%%%%%%%%%%%%%%%%%%%%%%%%%%%%%%%%%%%%%%%%%%%%%%%%%%%%%%%%%%%%%%%%

\section{Oscillation amplitude}  \label{sec.amp}

The amplitudes of individual modes are affected by the stochastic nature of
the excitation and by the (unknown) value of the mode lifetime.  To measure
the oscillation amplitude of \nuind{} in a way that is independent of these
effects, we have followed the method introduced by \citet{KBB2005}.  In
brief, this involves the following steps: (i)~smoothing the power spectrum
heavily to produce a single hump of excess power that is insensitive to the
fact that the oscillation spectrum has discrete peaks; (ii)~converting to
power density by multiplying by the effective length of the observing run
(4.42\,d, which we calculated from the area under the spectral window in
power); (iii)~fitting and subtracting the background noise; and
(vi)~multiplying by $\Delta\nu/3.0$ and taking the square root, in order to
convert to amplitude per oscillation mode.  For more details, see
\citet{KBB2005}.

\begin{figure*}
\epsscale{0.5}
\plotone{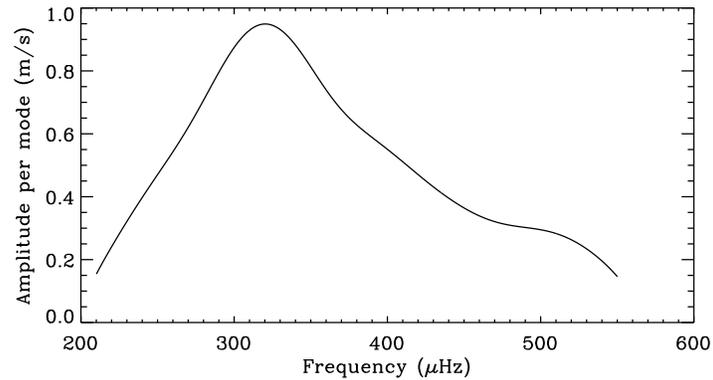}
\caption[]{\label{fig.ampsmooth} Smoothed oscillation spectrum of \nuind,
showing the amplitude per oscillation mode.  }
\end{figure*}

The result is shown in Fig.~\ref{fig.ampsmooth}.  The peak amplitude per
mode is 0.95\,\ms, which occurs at a frequency of $\nu_{\rm
max}=320$\,\muHz{} (period 52\,min).  This value of $\nu_{\rm max}$ is
consistent with that expected from scaling the acoustic cutoff frequency of
the Sun \citep{BGN91,K+B95}.  The observed peak amplitude is 4.6 times the
solar value, when the latter is measured using stellar techniques
\citep{KBB2005}, which is substantially less than the value of 7.3 expected
from the $L/M$ scaling proposed by \citet{K+B95} \new{but is in good
agreement with the $(L/M)^{0.7}$ scaling suggested for main-sequence stars
by \citet{SGA2005}.  A measurement of the mode lifetimes in \nuind{} would
be particularly useful.}

\section{Conclusions}

We have observed solar-like oscillations in the metal-poor subgiant star
\nuind{} and measured the large frequency separation.  We used this,
together with the location of the star in the H-R diagram and standard
evolutionary models, to place constraints on the stellar parameters.  Our
results, summarized in Table~\ref{tab.params}, confirm that \nuind{} has a
low mass and a large age and represent the first application of
asteroseismology to a metal-poor star.  Further constraints on the
parameters, particularly the mixing length, should come from comparing
individual oscillation frequencies with theoretical models.

\acknowledgments

We are extremely grateful to Conny Aerts for agreeing a time swap on
CORALIE that allowed us to observe at the optimum time of year.  We also
thank Geoff Marcy for useful advice and enthusiastic support.  This work
was supported financially by the Australian Research Council, the Swiss
National Science Foundation, the Danish Natural Science Research Council,
the Danish National Research Foundation through its establishment of the
Theoretical Astrophysics Center, and by a research associate fellowship
from Penn State University.  We further acknowledge support by NSF grant
AST-9988087 (RPB) and by SUN Microsystems.

\clearpage

\begin{table*}
\small
\caption{\label{tab.params} Parameters for \nuind{} (assuming $Z=0.001$
and $\alpha = 1.8\pm0.1$)}
\begin{center}
\begin{tabular}{lrclr}
\tableline
\tableline
\noalign{\smallskip}
$\Delta\nu~(\muHz)$          & 24.25 & $\pm$ & $0.25$     &  (1.0\%)\\
\Teff{} (K)                  & 5291  & $\pm$ & $34$       &   (0.64\%) \\
$M~(M_\sun)$                 & 0.847 & $\pm$ & $0.043$    &   (5.1\%) \\
Age (Gyr)                    & 11.4  & $\pm$ & $2.4$      &  (21\%) \\
$L~(L_\sun)$                 & 6.21  & $\pm$ & $0.23$     &   (3.7\%) \\
$R~(R_\sun)$                 & 2.97  & $\pm$ & $0.05$     &   (1.7\%) \\
$\log (g/\mbox{cm\,s}^{-2})$ & 3.421 & $\pm$ & $0.016$    &   (3.8\% in $g$) \\
angular diameter (mas)       & 0.956 & $\pm$ & $0.023$    &   (2.4\%) \\
\noalign{\smallskip}
\tableline
\end{tabular}
\end{center}
\end{table*}

\clearpage

\begin{thebibliography}{32}
\expandafter\ifx\csname natexlab\endcsname\relax\def\natexlab#1{#1}\fi
\expandafter\ifx\csname url\endcsname\relax
  \def\url#1{{\tt #1}}\fi

\bibitem[{Alonso} et~al.(1999){Alonso}, {Arribas}, \&
  {Mart{\'{\i}}nez-Roger}]{AAMR99}
{Alonso}, A., {Arribas}, S., \& {Mart{\'{\i}}nez-Roger}, C., 1999, A\&AS, 140,
  261.

\bibitem[Bedding et~al.(2004)Bedding, Kjeldsen, Butler, McCarthy, Marcy,
  O'Toole, Tinney, \& Wright]{BKB2004}
Bedding, T.~R., Kjeldsen, H., Butler, R.~P., McCarthy, C., Marcy, G.~W.,
  O'Toole, S.~J., Tinney, C.~G., \& Wright, J.~T., 2004, ApJ, 614, 380.

\bibitem[{Bouchy} et~al.(2005){Bouchy}, {Bazot}, {Santos}, {Vauclair}, \&
  {Sosnowska}]{BBS2005}
{Bouchy}, F., {Bazot}, M., {Santos}, N.~C., {Vauclair}, S., \& {Sosnowska}, D.,
  2005, A\&A, 440, 609.

\bibitem[Bouchy \& {Carrier}(2002)Bouchy, \& {Carrier}]{B+C2002}
Bouchy, F., \& {Carrier}, F., 2002, A\&A, 390, 205.

\bibitem[{Bouchy} et~al.(2001){Bouchy}, {Pepe}, \& {Queloz}]{BPQ2001}
{Bouchy}, F., {Pepe}, F., \& {Queloz}, D., 2001, A\&A, 374, 733.

\bibitem[Bretthorst(1988)]{Bre88}
Bretthorst, G.~L.
\newblock {\em Bayesian Spectrum Analysis and Parameter Estimation}, volume~48
  of {\em Lecture Notes in Statistics}.
\newblock Springer-Verlag: New York, 1988.

\bibitem[{Bretthorst}(2003)]{Bre2003}
{Bretthorst}, G.~L., 2003, In Willimas, C.~J., editor, {\em AIP Conf. Proc.
  659: Bayesian Inference and Maximum Entropy Methods in Science and
  Engineering}, page~3.
\newblock available from {\tt http://bayes.wustl.edu/}.

\bibitem[Brown et~al.(1991)Brown, Gilliland, Noyes, \& Ramsey]{BGN91}
Brown, T.~M., Gilliland, R.~L., Noyes, R.~W., \& Ramsey, L.~W., 1991, ApJ, 368,
  599.

\bibitem[Butler et~al.(2004)Butler, Bedding, Kjeldsen, McCarthy, O'Toole,
  Tinney, Marcy, \& Wright]{BBK2004}
Butler, R.~P., Bedding, T.~R., Kjeldsen, H., McCarthy, C., O'Toole, S.~J.,
  Tinney, C.~G., Marcy, G.~W., \& Wright, J.~T., 2004, ApJ, 600, L75.

\bibitem[Carrier \& {Bourban}(2003)Carrier, \& {Bourban}]{C+B2003}
Carrier, F., \& {Bourban}, G., 2003, A\&A, 406, L23.

\bibitem[{Carrier} et~al.(2005{\natexlab{a}}){Carrier}, {Eggenberger}, \&
  {Bouchy}]{CEB2005}
{Carrier}, F., {Eggenberger}, P., \& {Bouchy}, F., 2005{\natexlab{a}}, A\&A,
  434, 1085.

\bibitem[{Carrier} et~al.(2005{\natexlab{b}}){Carrier}, {Eggenberger},
  {D'Alessandro}, \& {Weber}]{CEDAl2005}
{Carrier}, F., {Eggenberger}, P., {D'Alessandro}, A., \& {Weber}, L.,
  2005{\natexlab{b}}, NewA, 10, 315.

\bibitem[Christensen-Dalsgaard(1982)]{ChD82}
Christensen-Dalsgaard, J., 1982, MNRAS, 199, 735.

\bibitem[Christensen-Dalsgaard(2004)]{ChD2004}
Christensen-Dalsgaard, J., 2004, Sol. Phys., 220, 137.

\bibitem[D'Antona et~al.(2005)D'Antona, {Cardini}, {Di Mauro}, {Maceroni},
  {Mazzitelli}, \& {Montalb{\'a}n}]{DACDiM2005}
D'Antona, F., {Cardini}, D., {Di Mauro}, M.~P., {Maceroni}, C., {Mazzitelli},
  I., \& {Montalb{\'a}n}, J., 2005, MNRAS, 363, 847.

\bibitem[{Gratton} et~al.(2000){Gratton}, {Sneden}, {Carretta}, \&
  {Bragaglia}]{GSC2000}
{Gratton}, R.~G., {Sneden}, C., {Carretta}, E., \& {Bragaglia}, A., 2000, A\&A,
  354, 169.

\bibitem[Gregory(2005)]{Gre2005}
Gregory, P.~C.
\newblock {\em Bayesian Logical Data Analysis for the Physical Sciences}.
\newblock Cambridge University Press, 2005.

\bibitem[{Guenther} et~al.(2005){Guenther}, {Kallinger}, {Reegen}, {Weiss},
  {Matthews}, {Kuschnig}, {Marchenko}, {Moffat}, {Rucinski}, {Sasselov}, \&
  {Walker}]{GKR2005}
{Guenther}, D.~B., {Kallinger}, T., {Reegen}, P., {Weiss}, W.~W., {Matthews},
  J.~M., {Kuschnig}, R., {Marchenko}, S., {Moffat}, A.~F.~J., {Rucinski},
  S.~M., {Sasselov}, D., \& {Walker}, G.~A.~H., 2005, ApJ.
\newblock in press (arXiv:astro-ph/0503695).

\bibitem[Hoffleit(1982)]{Hoffleit}
Hoffleit, D.
\newblock {\em The Bright Star Catalogue}.
\newblock Yale University Observatory, New Haven, 1982.

\bibitem[Kervella et~al.(2004)Kervella, {Th{\' e}venin}, {Morel}, {Berthomieu},
  {Bord{\' e}}, \& {Provost}]{KTM2004}
Kervella, P., {Th{\' e}venin}, F., {Morel}, P., {Berthomieu}, G., {Bord{\' e}},
  P., \& {Provost}, J., 2004, A\&A, 413, 251.

\bibitem[Kervella et~al.(2003)Kervella, {Th{\'e}venin}, {S{\'e}gransan},
  {Berthomieu}, {Lopez}, {Morel}, \& {Provost}]{KTS2003}
Kervella, P., {Th{\'e}venin}, F., {S{\'e}gransan}, D., {Berthomieu}, G.,
  {Lopez}, B., {Morel}, P., \& {Provost}, J., 2003, A\&A, 404, 1087.

\bibitem[Kjeldsen \& Bedding(1995)Kjeldsen, \& Bedding]{K+B95}
Kjeldsen, H., \& Bedding, T.~R., 1995, A\&A, 293, 87.

\bibitem[Kjeldsen et~al.(2003)Kjeldsen, Bedding, Baldry, Bruntt, Butler,
  Fischer, Frandsen, Gates, Grundahl, Lang, Marcy, Misch, \& Vogt]{KBB2003}
Kjeldsen, H., Bedding, T.~R., Baldry, I.~K., Bruntt, H., Butler, R.~P.,
  Fischer, D.~A., Frandsen, S., Gates, E.~L., Grundahl, F., Lang, K., Marcy,
  G.~W., Misch, A., \& Vogt, S.~S., 2003, AJ, 126, 1483.

\bibitem[Kjeldsen et~al.(2005)Kjeldsen, Bedding, Butler, Christensen-Dalsgaard,
  Kiss, McCarthy, Marcy, Tinney, \& Wright]{KBB2005}
Kjeldsen, H., Bedding, T.~R., Butler, R.~P., Christensen-Dalsgaard, J., Kiss,
  L., McCarthy, C., Marcy, G.~W., Tinney, C.~G., \& Wright, J.~T., 2005, ApJ,
  635, 1281.

\bibitem[{Lambert} \& {McWilliam}(1986){Lambert}, \& {McWilliam}]{L+McW86}
{Lambert}, D.~L., \& {McWilliam}, A., 1986, ApJ, 304, 436.

\bibitem[{Martic} et~al.(2004){Martic}, {Lebrun}, {Appourchaux}, \&
  {Schmitt}]{MLA2004b}
{Martic}, M., {Lebrun}, J.~C., {Appourchaux}, T., \& {Schmitt}, J., 2004, In
  {\em SOHO 14/GONG 2004 Workshop, Helio- and Asteroseismology: Towards a
  Golden Future}, ESA SP-559, page 563.
\newblock arXiv:astro-ph/0409126.

\bibitem[{Mosser} et~al.(2005){Mosser}, {Bouchy}, {Catala}, {Michel}, {Samadi},
  {Th{\' e}venin}, {Eggenberger}, {Sosnowska}, {Moutou}, \& {Baglin}]{MBC2005}
{Mosser}, B., {Bouchy}, F., {Catala}, C., {Michel}, E., {Samadi}, R., {Th{\'
  e}venin}, F., {Eggenberger}, P., {Sosnowska}, D., {Moutou}, C., \& {Baglin},
  A., 2005, A\&A, 431, L13.

\bibitem[{Nissen} et~al.(1997){Nissen}, {Hoeg}, \& {Schuster}]{NHS97}
{Nissen}, P.~E., {Hoeg}, E., \& {Schuster}, W.~J., 1997, In Battrick, B.,
  editor, {\em Hipparcos Venice '97 Symposium}, page 225. ESA SP-402.
\newblock {\tt\small
  http://\linebreak[0]astro.estec.esa.nl/\linebreak[0]SA-general/\linebreak[0]%
Projects/\linebreak[0]Hipparcos/\linebreak[0]venice.html}.

\bibitem[Pijpers et~al.(2003)Pijpers, {Teixeira}, {Garcia}, {Cunha},
  {Monteiro}, \& {Christensen-Dalsgaard}]{PTG2003}
Pijpers, F.~P., {Teixeira}, T.~C., {Garcia}, P.~J., {Cunha}, M.~S., {Monteiro},
  M.~J.~P.~F.~G., \& {Christensen-Dalsgaard}, J., 2003, A\&A, 406, L15.

\bibitem[{Samadi} et~al.(2005){Samadi}, {Goupil}, {Alecian}, {Baudin},
  {Georgobiani}, {Trampedach}, {Stein}, \& {Nordlund}]{SGA2005}
{Samadi}, R., {Goupil}, M.-J., {Alecian}, E., {Baudin}, F., {Georgobiani}, D.,
  {Trampedach}, R., {Stein}, R., \& {Nordlund}, {\AA}., 2005, JA\&A, 26, 171.

\bibitem[{Spergel} et~al.(2003){Spergel}, {Verde}, {Peiris}, {Komatsu},
  {Nolta}, {Bennett}, {Halpern}, {Hinshaw}, {Jarosik}, {Kogut}, {Limon},
  {Meyer}, {Page}, {Tucker}, {Weiland}, {Wollack}, \& {Wright}]{SVP2003}
{Spergel}, D.~N., {Verde}, L., {Peiris}, H.~V., {Komatsu}, E., {Nolta}, M.~R.,
  {Bennett}, C.~L., {Halpern}, M., {Hinshaw}, G., {Jarosik}, N., {Kogut}, A.,
  {Limon}, M., {Meyer}, S.~S., {Page}, L., {Tucker}, G.~S., {Weiland}, J.~L.,
  {Wollack}, E., \& {Wright}, E.~L., 2003, ApJS, 148, 175.

\bibitem[{Th{\'e}venin} et~al.(2005){Th{\'e}venin}, {Kervella}, {Pichon},
  {Morel}, {di Folco}, \& {Lebreton}]{TKP2005}
{Th{\'e}venin}, F., {Kervella}, P., {Pichon}, B., {Morel}, P., {di Folco}, E.,
  \& {Lebreton}, Y., 2005, A\&A, 436, 253.

\end{thebibliography}
\end{document}